\documentclass[11pt]{article}
\usepackage{hyperref}
\usepackage{amssymb}
\usepackage{graphicx}
\usepackage{amsmath}
\usepackage{physics}
\usepackage{authblk}
\newcommand{\bea}{\begin{eqnarray}}
\newcommand{\eea}{\end{eqnarray}}
\newcommand{\be}{\begin{equation}}
\newcommand{\ee}{\end{equation}}
\usepackage[section]{placeins}
\usepackage[a4paper, total={6.8in, 10in}]{geometry}
\numberwithin{equation}{section}
\title{Time-space noncommutativity and Casimir effect}
\author{E. Harikumar \thanks{harisp@uohyd.ernet.in}, Suman Kumar Panja
\thanks{sumanpanja19@gmail.com} and Vishnu Rajagopal \thanks{vishnurajagopal.anayath@gmail.com}}
\affil{School of Physics, University of Hyderabad, \\Central University P.O, Hyderabad-500046, Telangana, India}
\date{}
\begin{document}

\maketitle
\begin{abstract}
We show that the Casimir force and energy are modified in the $\kappa$-deformed space-time. This is shown by 
solving the Green's function corresponding to $\kappa$-deformed scalar field equation in presence of two parallel 
plates, modelled by $\delta$-function potentials. Exploiting the relation between Energy-Momentum tensor and Green's 
function, we calculate corrections to Casimir force, valid up to second order in the deformation parameter. The 
Casimir force is shown to get corrections which scale as $L^{-4}$ and $L^{-6}$ and both these types of corrections 
produce attractive forces. Using the measured value of Casimir force, we show that the deformation parameter should 
be below $10^{-23}$m.
\end{abstract}

\section{Introduction}
The nature of gravitational force is not yet known at microscopic scales irrespective of the fact that it is one of the oldest known forces. Attempts to understand working of gravity at extremely short distance scales are naturally one of the most active areas of research. Various approaches, based on different paradigms are being pursued for many years to understand quantum theory of gravity. String theory, loop gravity, causal dynamical triangulations, emergent gravity, asymptotically safe models, non-commutative geometry\cite{seiberg-witten,Glikman,dop,am,madore,bal,connes} etc are some of these approaches. Many of these methods have some common features and introduction of a fundamental length scale below which quantum effects of gravity is prominent, is one of them\cite{Glikman,dop}. Thus incorporating a fundamental length scale in models of quantum gravity is essential and non-commutative geometry provides an elegant way to do this\cite{Glikman,dop}. Non-commutative geometry also appears in low energy limits of string theory, loop gravity etc. 

Two different types of non-commutative space-times\cite{connes} that have been studied extensively are Moyal 
space-time\cite{nek} and $\kappa$-deformed space-time\cite{luke}. In the former, the commutator between space-time 
coordinates are constants while in the later the commutator between time and space coordinates are proportional to 
the space coordinates themselves. Field theory models on these space-time have been constructed and studied with various motivations\cite{nek}. 

Existence of a fundamental length scale in models of microscopic theory of gravity necessitated the modification of 
the principle of relativity and this lead to the formulation of deformed special relativity\cite{dsr}. 
The space-time associated with deformed special relativity is shown to be the 
$\kappa$-deformed space-time\cite{majid}.  The inherent features of quantum theories on non-commutative space-times are non-locality, non-linearity and introduction of a length parameter. There have been many studies to analyse the effect of this fundamental length scale and its possible signals\cite{jabbari,sdas,posp,gupta,siva,kapoor,ravi,hari,btz,vish}. It is also known that the symmetries of these models are realised as Hopf algebras\cite{Chaichan, Wess,luk,hopf} and their quantisation rules are non-trivial. Thus it is of intrinsic interest to study the implications of minimal length (introduced through the space-time non-commutativity) and that of modified quantisation rules  in physical phenomena. Introduction of minimal length scale also leads to modification of commutation relations between coordinates and momenta and this results in generalised uncertainty principle(GUP). Non-commutativity is also known to result changes in the energy-momentum relations. Various implications of these modifications have been studied\cite{jabbari,sdas,posp,gupta,siva,kapoor,ravi,hari,btz,vish} . 

One of the well studied phenomena where the length scale plays an important role is the Casimir effect\cite{kam}. It has been known that two conducting plates kept parallel, separated by very small distance, do attract each other. This force is known to arise due to the vacuum fluctuations of the quantised fields. This phenomenon has been  studied for various other geometrical configurations such as conducting sphere and plate etc\cite{kam}.  

Casimir force has been investigated experimentally  with great accuracy and at present it has been measured for separations of the plates of the order of few micrometers\cite{robert,brax,sed}. These results have been used to obtain constraints on the corrections to Newtonian gravity\cite{1306.4979}.

Though the length scales at which measurements of Casimir effect are made at present and quantum gravity effects are expected to be important are very different, it is worth studying the possible modification of Casimir effect due to space-time non-commutativity. Many results in this direction have been reported in recent times. Effects of minimal length, GUP and modified dispersion relations on Casimir force and corresponding energy have been investigated by various authors\cite{kh.nou,kempf,0707.0405,1112.2924,SB,0704.2251,0711.4272,sp,npb127,jab}.

Generalised commutation relations between coordinates and momenta as well as those between coordinates among themselves have been studied in the context of quantum gravity and they are known to be related to generalised uncertainty relations and minimal length scale. These coordinates are realised in terms of usual coordinates, momenta and parameter(s) characterising the modified commutation relations. Further, it was shown that the maximally localised state corresponding to these operators are different from the plane waves, yet they are normalisable\cite{kh.nou, kempf}. The GUPs have been shown to lead to non-trivial measure in  momentum space. The modification brought in by these changes to different physical models have been analysed. In \cite{kh.nou}, the correction to Casimir effect between parallel plates was analysed.  In this work, expanding the quantised Maxwell field in terms of maximally localised state, it was shown that the commutation relations between the creation and annihilation operators are modified. This results in the change in the energy spectrum  and in particular, changes the zero-point energy. This zero-point energy was used to find the correction to Casimir energy and force. This correction due to minimal length can be attractive or repulsive, depending on the value of the minimal length\cite{kh.nou}. Note that, in\cite{kh.nou} it was shown that the zero-point energy of free Maxwell field itself is modified due to the minimal length scale. The entire effect of minimal length is introduced through the maximally localised states, which in turn, result in the modified oscillator algebra.

In\cite{0707.0405}, for a specific GUP, for the case of Casimir-Polder interaction, using the approach of \cite{kh.nou, kempf}, force between parallel plates was calculated. Treating the correction to interaction potential between  neutral atoms(of the plates), i.e., between the electric dipole moments of these atoms with electric field, perturbatively, correction to Casimir energy was calculated. This was done for atoms/molecules separated by short/large distances. The modification in the force between parallel plates were calculated by these authors. The corrections in 3-dimensional space were seen to be scaling as $L^{-7}$ and $L^{-9}$, where $L$ is the separation between the plates, respectively. Here the correction terms all were seen to be attractive for both short/long distances. This method of using maximally localised states have been applied to many different possible GUPs and the Casimir effect for these cases were calculated in\cite{1112.2924}. Here too, correction was found to be attractive. In this case, the correction due to minimal length was shown to be scaled as $L^{-5}$. For generic dispersion relations which are analytical, the correction to Casimir force was calculated in\cite{SB}. The dependence of Casimir force for these dispersion relations on minimal length was also investigated in \cite{SB}.

Casimir force between two parallel plates in the Moyal space-time was studied in \cite{0704.2251, 0711.4272}. The coordinates of Moyal space-time satisfies 
\be
[x^\mu, x^\nu]=i\theta^{\mu\nu}
\ee
where $\theta^{\mu\nu}$ is a constant Lorentz tensor. The non-commutativity of the space-time introduces difficulties in implementing boundary conditions as the space-time points do not have any well defined meaning in non-commutative space-time. Imposing boundary conditions play a very important role in the calculation of Casimir effect and thus its calculation in non-commutative space-time has to be done with care. In \cite{0704.2251}, for even dimensional Moyal space-time, using a coherent state approach\cite{sp} and smeared boundary condition, Casimir force between two parallel plates was calculated. In the coherent state approach, the non-commutative plane waves are mapped to the usual plane waves with an additional damping factor which depends quadratically on momenta.

It is known that the quadratic part of the action in Moyal space-time is same as that in the commutative space-time\cite{nek} and the effects of non-commutativity appear only through the interactions. Thus the vacuum fluctuations relevant for Casimir effect is independent of non-commutative parameter in the Moyal space-time\cite{0711.4272}. But for Casimir effect with circular boundary conditions, the interaction term is non-trivial and  Casimir effect is modified in Moyal space-time and this was calculated in\cite{0711.4272}.

In \cite{npb127}, Casimir effect in the $\kappa$-space-time was calculated using a specific form of $\kappa$-deformed scalar theory\cite{luk}. $\kappa$-space-time is one where the commutator of time coordinate with space coordinates are proportional to space coordinates, while the commutators between space coordinates vanish, i.e., they obey
\be
[x^0, x^i]=ax^i,~~~[x^i, x^j]=0,~~~~a=\frac{1}{\kappa}.
\ee
The $\kappa$-deformed Lagrangian has derivatives to all higher orders up to infinity. Thus definition of conjugate momentum is not straight forward as in the commutative  theory. In the approach used in \cite{npb127}, the conjugate momentum used is not unique. The issue is addressed by calculating the expression for energy corresponding to the scalar theory without using an explicit form of conjugate momentum. This is achieved by modifying the integration measure in the momentum space. This modification allows one to have the same commutation relations between the creation and annihilation operators appearing in the quantised field, but the equal time commutator between fields is now become non-vanishing. Using thus obtained quantised energy of the field and treating deformed electromagnetic theory to be equivalent to scalar theory with two polarisations, vacuum energy of the $\kappa$-deformed Maxwell field is calculated. Using this, Casimir energy is calculated where the boundary conditions imposed are the same as the ones in commutative space-time. Casimir energy is shown to scale as $L^{-3}$ in this case\cite{npb127}.

For $1+1$ dimension, in commutative case, Casimir effect for two parallel plates, has been rederived by considering two $\delta$-function potentials \cite{leip}, which  imply that the  parallel plates are at $x=0$ and $x=L$, respectively. Casimir Force has been calculated on plate at $x=L$, by using discontinuity of stress tensor. For the  massless case, $xx$ component of stress tensor has been calculated at a point, just left of the plate at $x=L$ i.e $\hat{T}^{xx}\bigg| _{x=L^-}$ and another point, just of the right of the plate at $x=L$ i.e $\hat{T}^{xx}\bigg| _{x=L^+}$. Then the force acting on the plate at $x=L$, due to quantum fluctuations, has been derived by taking the difference between vacuum expectation value of these  stress tensors using strong coupling of $\delta$-function potentials as,
\begin{equation}
    \hat{F}=<\hat{T}^{xx}>\bigg| _{x=L^-}-<\hat{T}^{xx}>\bigg|_{x=L^+}.
\end{equation}
The difference in the vacuum expectation value of $\hat{T}^{\alpha\mu}_{xx'}$ at $x$ and $x'$ can be expressed in terms of time ordered product of  fields at $x$ and $x'$ as,
\begin{equation}
 \hat{T}^{\alpha\mu}_{xx'}=\hat{O}^{\alpha\mu}(\partial,\partial')T(\phi(x),\phi(x')),
\end{equation}
where $\hat{O}^{\alpha\mu}(\partial,\partial')$ stands for a combination of operators, which acts on time ordered product of fields at nearby points.
Since time ordered product of fields at $x$ and $x'$ is given by Green's function, the eqn(1.3) becomes
\begin{equation}
  \hat{F}=-i\hat{O}^{xx'}(\partial,\partial')G(x,x')
\end{equation}
To calculate $G(x,x')$ in different regions, Euler-Lagrangian equation of scalar field theory with interaction part  is solved with  Dirichlet boundary condition. Thus boundary condition plays a significant role in the calculation of Casimir Force. Then Casimir energy is calculated by integrating the Casimir force over the length between two parallel plates, taken to be infinitely separated, i.e.,
\begin{equation}
 \hat{E}_{\lambda,\lambda'\rightarrow \infty}=-\int_0^{\infty} dx\hat{F}_{\lambda,\lambda'\rightarrow\infty}.
\end{equation}
Here $\lambda$ and $\lambda'$ are coupling constants, $\lambda,\lambda'\rightarrow\infty$ indicate the strong coupling limit.

In this paper, we calculate the Casimir force and energy between two parallel plates by analysing the $\kappa$-deformed Klein-Gordon theory. In our approach, we start with the Klein-Gordon Lagrangian which is invariant under the action of undeformed $\kappa$-Poincare algebra\cite{hopf}. This Lagrangian is written in terms of the field defined in the commutative space-time and this allows us to use calculational tools of usual field theory in deriving effect of vacuum fluctuations. After calculating the energy-momentum tensor corresponding to $\kappa$-deformed scalar theory in $1+1$ dimensions, we derive the corresponding Casimir force and energy.

We use the same boundary conditions as commutative case as $\kappa$-deformed Klein-Gordon equation is written in commutative space-time. Using these boundary conditions and $\kappa$-deformed Euler-Lagrangian equation for scalar field theory, we have calculated Green's function solution valid up to order $a^2$, in different regions. Using equation (1.5), we have calculated time ordered product of fields at nearby points. Then using higher order derivative theory, we have calculated  energy momentum tensor $\hat{T}^{\alpha\mu}$, with dependence on $\kappa$-deformation parameter, up to first non vanishing term i.e., $a^2$, from $\kappa$-deformed Lagrangian. We have symmetrised Energy-Momentum expression and obtained vacuum expectation value of Energy-Momentum tensor in the form,
\begin{equation}
 \hat{T}^{\alpha\mu}_{xx'}=\hat{O}^{\alpha\mu}(\partial,\partial')T(\phi(x),\phi(x')).
\end{equation}
Here $\hat{O}^{\alpha\mu}(\partial,\partial')$ and $T(\phi(x),\phi(x'))$ both depend on $\kappa$-deformation parameter $a$. Using the above relation, in the same way as in the commutative case, we have derived difference of stress tensor at the left of the plate and right of the plate at, $x=L$ and calculated the Casimir force. 

This paper is organised as follows. In the next section, we construct $\kappa$-deformed Lagrangian describing scalar field theory in the presence of two parallel plates. Here we first give a brief summary of the $\kappa$-deformed scalar theory \cite{trg} written using the quadratic Casimir of the undeformed $\kappa$-Poincare algebra \cite{hopf}. This Lagrangian, describing the interaction of $\kappa$-scalar field with parallel plates is written in terms of commutative variables and valid up to first non-vanishing terms in the deformation parameter $a$. This allows us to impose the boundary conditions required to calculate Casimir force and energy as in the commutative space-time. The new terms in the Lagrangian due to $\kappa$-deformation are all of higher derivatives. Our main results are obtained in Sec.3. Here first we discuss the solution to the Euler-Lagrangian equation, describing the interaction of the scalar field with parallel plates in $\kappa$-space-time. This is obtained perturbatively and valid up to second order in $a$. Then we give the reduced Green's function for different regions of interest and present the well known relation between the Green's function and the vacuum expectation value of the energy-momentum tensor (in the commutative space-time). We then derive the energy-momentum tensor corresponding to the $\kappa$-deformed scalar theory, constructed in Sec.2. This energy-momentum tensor is derived using the approach of higher derivative theories. We then discuss the expression for vacuum expectation value of the $\kappa$-deformed energy-momentum tensor in terms of the Green's function. Using the reduced Green's function solutions obtained in different regions, we then explicitly calculate the Casimir energy and Casimir force for $1+1$ dimension and generalise these results to $3+1$ dimensions. Comparing with experimental results on Casimir effect, we show that the deformation parameter has an upper bound of $10^{-23}$. Our concluding remarks are given in Sec.4. In appendix, we give the essential details of the construction of energy-momentum tensor for our higher derivative Lagrangian. Here we work with $\eta_{\mu\nu}=diag(-1,+1,+1,+1)$. 
\section{$\kappa$-deformed Lagrangian for scalar field }
The coordinates of $\kappa$-deformed space-time obey the Lie-algebra type commutation relations given by
\begin{equation}\label{def1}
[\hat{x}_i,\hat{x}_j]=0,~~[\hat{x}_0,\hat{x}_i]=ia\hat{x}_i,
\end{equation}
where the deformation in parameter $a$ has the dimension of length.

In the approach we take here, the coordinates of $\kappa$-deformed space-times are re-expressed in terms of commutative coordinates and their derivatives \cite{hopf} as
\begin{equation}\label{def2}
 \hat{x}_i=x_i\varphi(A),~~\hat{x}_0=x_0,
\end{equation}
with $A=ap_0$. The consistency of Eq.(\ref{def1}) and Eq.(\ref{def2}) implies that $\varphi(A)$ satisfy the 
conditions
\begin{equation}
 \varphi'=-\varphi,{\rm~and}~~~\varphi(0)=1.
\end{equation}
It is known that different choices for $\varphi(A)$, satisfying these requirements are possible \cite{hopf}.

Following \cite{hopf}, we define the generators of the symmetry algebra of $\kappa$-deformed space-time by the relations
\begin{equation}\label{up1}
 [D_{\mu},D_{\nu}]=0,
\end{equation}
\begin{equation}\label{up2}
 [M_{\mu\nu},D_{\lambda}]=\eta_{\nu\lambda}D_{\mu}-\eta_{\mu\lambda}D_{\nu},
\end{equation}
\begin{equation}\label{up3}
 {\rm~and~}~~[M_{\mu\nu},M_{\lambda\rho}]=\eta_{\mu\rho}M_{\nu\lambda}+\eta_{\nu\lambda}M_{\mu\rho}-\eta_{\nu\rho}M_{\mu\lambda}-\eta_{\mu\lambda}M_{\nu\rho}
\end{equation}
where $D_{\mu}$  is Dirac derivative, which transform as a four vector and its explicit form is given by
\begin{equation}\label{dirac}
D_{i}=\partial_{i}\frac{e^{-A}}{\varphi},~~
D_{0}=\partial_{0}\frac{\textnormal{sinh}A}{A}-ia\partial_{i} ^2 \frac{e^{-A}}{2\varphi^2}.
\end{equation}
The generator $M_{\mu\nu}$ is also deformed and expressed in terms of coordinates of the commutative space-time, 
their derivatives and the deformation parameter as
\begin{equation}
 M_{\mu\nu}=(\hat{x}_{\mu}D_{\nu}-\hat{x}_{\nu}D_{\mu})Z,
\end{equation}
where $Z^{-1}=iaD_0+\sqrt{1+a^2D^2}$.

Note that the defining relations, Eq.(\ref{up1}), Eq.(\ref{up2}) and Eq.(\ref{up3}) of the symmetry algebra 
are the same as that of the Poincare algebra, but the generators are deformed. This algebra has been referred as 
undeformed $\kappa$-Poincare algebra \cite{trg,ravi}. This should be contrasted with the $\kappa$-Poincare algebra,
where the defining relations are deformed\cite{luke,majid}.

The  corresponding quadratic Casimir of the undeformed $\kappa$-Poincare algebra is $D_{\mu}D^{\mu}$ and given by
\begin{equation}
D_{\mu}D^{\mu}=\Box\Big(1+\frac{a^2}{4}\Box\Big),
\end{equation}
where $\Box$ is given as
\begin{equation}
\Box=\partial_{i} ^2 \frac{e^{-A}}{\varphi^2}+2\partial_{0} ^2\frac{(1-\textnormal{cosh}A)}{A^2},
\end{equation}
The corresponding $\kappa$-deformed dispersion relation is,
\begin{equation}
  -p_i^2+\frac{4}{a^2}{\textnormal{sinh}^2\Big(\frac{ap_0}{2}\Big)}+\frac{a^2}{4}\Big[-p_i^2+4\textnormal{sinh}^2\Big(\frac{ap_0}{2}\Big)\Big]^2=0
\end{equation}
Thus the generalized massless Klein-Gordon equation on $\kappa$-deformed space-time takes the form
\begin{equation}
 \Box \Big(1+\frac{a^2}{4}\Box\Big)\phi(x)=0.
\end{equation}
Now we write down the $\kappa$-deformed real scalar field Lagrangian.
 In terms of Dirac derivatives, Lagrangian is
\begin{equation}\label{kg}
    \mathcal{L}_0=-\frac{1}{2}\phi D_{\mu} D^{\mu}\phi.
\end{equation}
By choosing $\varphi$, we can obtain different realisations of $\kappa$-deformed Lagrangian. In this paper we have chosen $\varphi=e^{-\frac{A}{2}}$. To study Casimir effect between two parallel plates, one introduces these plates through their interaction with fields. This interaction of the two parallel plates kept in the vacuum at $x=0$ and $x=L$ is described by an interaction Lagrangian, which is given as
\begin{equation}\label{int}
 \mathcal{L_{\textnormal{int}}}=-{\frac{\lambda}{2L}\phi^2\delta(x)}-{\frac{\lambda^\prime}{2L}}\phi^2\delta(x-L).
\end{equation}
Thus the total Lagrangian is written as
\begin{equation}
   \mathcal{L} =\mathcal{L}_0+\mathcal{L_{\textnormal{int}}}=-\frac{1}{2}\phi D_{0}D^{0}\phi-\frac{1}{2}\phi D_{i}D^{i}\phi+\mathcal{L_{\textnormal{int}}}.
\end{equation}
Using Dirac derivative given in Eq.(\ref{dirac}) we expand the above Lagrangian and keep up to first non vanishing terms in $a$. Here we consider the  deformed Lagrangian in $1+1$ dimension given by,
\begin{equation}\label{lag}
  \mathcal{L}=\frac{1}{2}{\partial_m\phi\partial_m\phi}-\frac{1}{2}{\partial_0\phi\partial_0\phi}-\frac{a^2}{8}{\partial_m\phi\partial_m\partial_0\partial_0\phi}+\frac{a^2}{8}{\partial_m\partial_0\phi\partial_m\partial_0\phi}+\frac{a^2}{6}{\partial_0\phi\partial_0\partial_0\partial_0\phi}-\frac{a^2}{8}{\partial_m\partial_m\phi\partial_m\partial_m\phi}+\mathcal{L_{\textnormal{int}}}.
\end{equation}
Note that the interaction part is same as that in the commutative case and is given in eq.(2.8). This is possible since Lagrangian in Eq.(\ref{lag}) is written in terms of commutative variables. Thus by using $\kappa$-deformed  Lagrangian written in terms of commutative variables, we avoid issues related to implementation of boundary condition in non-commutative space-time.
\section{$\kappa$-deformed Casimir effect}
In this section we obtain the vacuum expectation value of the deformed stress tensor, valid up to first nonvanishing term in $a$ i.e $a^2$. Using this, we obtain the deformed Casimir force and deformed Casimir energy valid up to  $a^2$, in both $1+1$ and $3+1$ dimensions, separately.

The Euler-Lagrangian equation following from the above Lagrangian in Eq.(\ref{lag}) is given by
\begin{equation}\label{ce}
 \Big(\partial_i^2-\partial_0^2+\frac{a^2}{3}\partial_0^4+\frac{a^2}{4}\partial_i^4-\frac{a^2}{2}\partial_i^2\partial_0^2+\frac{\lambda}{L}\delta(x)+\frac{\lambda'}{L}\delta(x-L)\Big)\phi(x)=0.
\end{equation}
We solve this equation using perturbative method and thus we assume the solution $\hat{\phi}(x)$ to be of the form 
\begin{equation}\label{per}
  \hat{\phi}(x)=\phi_0(x)+a\alpha\phi_1(x)+a^2\beta\phi_2(x),
\end{equation}
where the dimensions of $\alpha$ and $\beta$ are $\frac{1}{L}$ and $\frac{1}{L^2}$, respectively. Substituting Eq.(\ref{per}) in Eq.(\ref{ce}), we get three equations,
\begin{equation}\label{a0}
  \Big(\partial_i^2-\partial_0^2+\frac{\lambda}{L}\delta(x)+\frac{\lambda'}{L}\delta(x-L)\Big)\phi_0(x)=0,
\end{equation}
\begin{equation}\label{a1}
  \Big(\partial_i^2-\partial_0^2+\frac{\lambda}{L}\delta(x)+\frac{\lambda'}{L}\delta(x-L)\Big)\phi_1(x)=0,
\end{equation}
\begin{equation}\label{a2}
  \beta\Big(\partial_i^2-\partial_0^2+\frac{\lambda}{L}\delta(x)+\frac{\lambda'}{L}\delta(x-L)\Big)\phi_2(x)+\Big(\frac{\partial_0^4}{3}+\frac{\partial_i^4}{4}-\frac{\partial_i^2\partial_0^2}{2}\Big)\phi_0(x)=0,
\end{equation}
respectively. From these equations we notice that $\phi_1(x)=\phi_0(x)$. In the limit $\lambda,\lambda'\rightarrow 0$, we have $\phi(0)=\phi(L)=0$ and thus get $\phi_0(x)=C(e^{ik'x}-e^{-ik'x})=C(e^{ikx}-e^{-ikx})e^{-iwt}$, where  $C$ is constant, $k'$ is four-vector momentum, wave vector $k=\frac{n\pi}{L}$,where $n$ is the number of mode in between parallel plates and $w$ is the wave frequency. Using these, we solve Eq.(\ref{a2}) and this gives $\phi_2(x)=\frac{13}{24\beta}a^2k^2\phi_0(x)$, where $k$ is wave vector whose dimension is $\frac{1}{L}$.Thus the solution to Euler-Lagrangian equation, valid up to second order in $a$ is
\begin{equation}\label{p1}
    \hat{\phi}(x)=(1+a\alpha+a^2\frac{13}{24}k^2)\phi_0(x).
\end{equation}
Note that the above solution has correction terms due to non-commutativity, valid up to order $a^2$.

Since the Casimir force is calculated using Green's function, first we obtain the Green's function from the relation
\begin{equation}
  G(x,x')=i<T\big(\phi(x)\phi(x')\big)>.
\end{equation}          
The Fourier transform of Green's function is expressed as,
\begin{equation}
  G(x,x')=\int \frac{dw}{2\pi}e^{-iw(t-t')}g(x,x',w).
\end{equation}          
and the reduced Green's function, $g(x,x',w)$ obeys the equation,
\begin{equation}
 \Big({\frac{\partial^2}{\partial x^2}}-\frac{\partial^2}{\partial t^2}+\frac{\lambda}{L}\delta(x)+\frac{\lambda'}{L}\delta(x-L)\Big)g(x,x',w)=-\delta(x-x').
\end{equation}
Using $k^2=-w^2$, above equation becomes
\begin{equation}
 -\Big(\frac{\partial^2}{\partial x^2}+k^2+\frac{\lambda}{L}\delta(x)+\frac{\lambda'}{L}\delta(x-L)\Big)g(x,x',k)=\delta(x-x') 
\end{equation} 
Now we find the solution for reduced Green's function in different regions.
Thus the reduced Green's function in the three regions of interest are \cite{mil}
\begin{eqnarray}\label{reg1}
  g(x,x')&=&\frac{1}{2k}e^{-k|x-x'|}+\frac{1}{2k\Delta}\frac{\lambda\lambda'}{(2kL)^2}2\textnormal{cosh}(k|x-x'|)-\frac{1}{2k\Delta}{\frac{\lambda}{2kL}\Big(1+\frac{\lambda'}{2kL}\Big)e^{2kL}e^{-k(x+x')}}\nonumber\\& &-\frac{1}{2k\Delta}{\frac{\lambda'}{2kL}\Big(1+\frac{\lambda}{2kL}\Big)e^{k(x+x')}};\textnormal{ where, }0<x,x'<L
\end{eqnarray}
\begin{equation}\label{reg2}
  g(x,x')=\frac{1}{2k}e^{-k|x-x'|}+{\frac{1}{2k\Delta}e^{-k(x+x'-2L)}}\bigg({-{\frac{\lambda}{2kL}\Big(1-\frac{\lambda'}{2kL}\Big)}}{-{\frac{\lambda'}{2kL}\Big(1+\frac{\lambda}{2kL}}\Big)e^{2kL}}\bigg);\textnormal{ where, }L<x,x'  
\end{equation}
\begin{equation}\label{reg3}
  g(x,x')=\frac{1}{2k}e^{-k|x-x'|}+{\frac{1}{2k\Delta}e^{k(x+x')}}\bigg({-{\frac{\lambda'}{2kL}\Big(1-\frac{\lambda}{2kL}\Big)}{-{\frac{\lambda}{2kL}}\Big(1+\frac{\lambda'}{2kL}\Big)e^{2kL}}}\bigg);\textnormal{ where, }x,x'<0    
\end{equation}
where
\begin{equation}\label{del}
   \Delta=\Big(1+\frac{\lambda}{2kL}\Big)\Big(1+\frac{\lambda'}{2kL}\Big)e^{2kL}-\frac{\lambda\lambda'}{(2kL)^2}.
\end{equation}
For a real scalar field in the commutative regime, vacuum expectation value of the stress tensor is calculated from Green's function. Using this the Casimir energy is determined.

For a free scalar field theory, the stress tensor is given as,
\begin{equation}
  T^{\mu\lambda}=\partial^{\mu}\phi(x)\partial'^{\lambda}\phi(x')-\frac{1}{2}\eta^{\mu\lambda}\partial_{\alpha}\phi(x)\partial'^{\alpha}\phi(x')
\end{equation}
which is written using time ordered product as
\begin{equation}
 T^{\mu\nu}_{x,x'}=\frac{1}{2}\Big(\partial^{\mu}\partial'^{\nu}+\partial^{\nu}\partial'^{\mu}-\eta^{\mu\nu}\partial^{\lambda}\partial'_{\lambda}\Big) T(\phi(x),\phi(x')).
\end{equation}
Thus the vacuum expectation value of $T^{\mu\lambda}$ is
\begin{equation}
  <T^{\mu\lambda}>=\frac{1}{2}\Big(\partial^{\mu}\partial'^{\lambda}+\partial^{\lambda}\partial'^{\mu}-\eta^{\mu\lambda}\partial^{\alpha}\partial'_{\alpha}\Big) <T(\phi(x),\phi(x'))>
\end{equation}
\begin{equation}
=\frac{1}{2}\Big(\partial^{\mu}\partial'^{\lambda}+\partial^{\lambda}\partial'^{\mu}-\eta^{\mu\lambda}\partial^{\alpha}\partial'_{\alpha}\Big)\frac{1}{i}G(x,x')
\end{equation}
Generalising this approach, we evaluate the vacuum expectation value for $\kappa$-deformed Energy-Momentum tensor. For this we first derive expression of Energy-Momentum tensor $\hat{T}^{\alpha\mu}$ in the $\kappa$-deformed space-time.
\subsection{Modified Energy Momentum tensor}
In this subsection we evaluate the deformed stress tensor from the Lagrangian describing the $\kappa$-deformed scalar field using the higher derivative formalism.
 
From higher derivative theory, the expression for the stress tensor is given (see appendix) as\cite{bollini}
\begin{eqnarray}
  \hat{T}^{\alpha\mu}&=&\Big(\frac{\partial \mathcal{L}}{\partial(\partial_\alpha \phi)}-\partial_\beta\frac{\partial \mathcal{L}}{\partial(\partial_\alpha\partial_\beta \phi)}+\partial_\beta\partial_\gamma\frac{\partial \mathcal{L}}{\partial(\partial_\alpha\partial_\beta \partial_\gamma\phi)}\Big)\partial^\mu\phi+\Big(\frac{\partial \mathcal{L}}{\partial(\partial_\alpha\partial_\beta \phi)}-\partial_\gamma\frac{\partial \mathcal{L}}{\partial(\partial_\alpha\partial_\beta \partial_\gamma\phi)}\Big)\partial_\beta\partial^\mu\phi\nonumber\\& &+\Big(\frac{\partial \mathcal{L}}{\partial(\partial_\alpha\partial_\beta \partial_\gamma\phi)}\Big)\partial_\beta\partial_\gamma\partial^\mu\phi-\eta^{\alpha\mu}\mathcal{L}.
\end{eqnarray}
Using this, we calculate the stress tensor for the Lagrangian given in Eq.(\ref{lag}). Consider the first term in 
the RHS of above equation, which we calculate explicitly as 
\begin{eqnarray}
 \frac{\partial\mathcal{L}}{\partial(\partial_{\alpha}\phi)}
 =\delta_0^{\alpha}\frac{\partial\mathcal{L}}{\partial(\partial_{0}\phi)}+\delta^{\alpha}_i\frac{\partial\mathcal{L}}{\partial(\partial_{i}\phi)}
=\partial^{\alpha}\phi+\frac{a^2}{6}\delta^{\alpha}_{0}\partial_0\partial_0\partial_0\phi-\frac{a^2}{8}\delta^{\alpha}_i\partial_i\partial_0\partial_0\phi.
\end{eqnarray}
Next we calculate the second term in the RHS, i.e
\begin{eqnarray}
 \partial_{\beta}\frac{\partial\mathcal{L}}{\partial(\partial_{\alpha}\partial_{\beta}\phi)}&=&\delta_0^{\alpha}\partial_0\frac{\partial\mathcal{L}}{\partial(\partial_{0}\partial_0\phi)}+\delta_i^{\alpha}\partial_0\frac{\partial\mathcal{L}}{\partial(\partial_i\partial_{0}\phi)}+\delta_0^{\alpha}\partial_i\frac{\partial\mathcal{L}}{\partial(\partial_{0}\partial_i\phi)}+\delta_i^{\alpha}\partial_j\frac{\partial\mathcal{L}}{\partial(\partial_{i}\partial_j\phi)}\nonumber\\ 
&=&\frac{a^2}{4}\delta^{\alpha}_i\partial_0\partial_i\partial_0\phi+\frac{a^2}{4}\delta^{\alpha}_0\partial_j\partial_j\partial_0\phi-\frac{a^2}{4}\delta^{\alpha}_i\partial_i\partial^j\partial_j\phi.
\end{eqnarray}
The third term is given by
\begin{eqnarray}
 \partial_{\beta}\partial_{\gamma}\frac{\partial\mathcal{L}}{\partial(\partial_{\alpha}\partial_{\beta}\partial_{\gamma}\phi)}&=&\delta^{\alpha}_0\partial_0\partial_0\frac{\partial\mathcal{L}}{\partial(\partial_{0}\partial_0\partial_0\phi)}+\delta^{\alpha}_i\partial_0\partial_0\frac{\partial\mathcal{L}}{\partial(\partial_{i}\partial_0\partial_0\phi)}+\delta^{\alpha}_0\partial_i\partial_0\frac{\partial\mathcal{L}}{\partial(\partial_{0}\partial_i\partial_0\phi)}+\delta^{\alpha}_i\partial_j\partial_0\frac{\partial\mathcal{L}}{\partial(\partial_{i}\partial_j\partial_0\phi)}\nonumber\\& &+\delta^{\alpha}_0\partial_0\partial_i\frac{\partial\mathcal{L}}{\partial(\partial_{0}\partial_0\partial_i\phi)}+\delta^{\alpha}_i\partial_0\partial_j\frac{\partial\mathcal{L}}{\partial(\partial_{i}\partial_0\partial_j\phi)}+\delta^{\alpha}_0\partial_i\partial_j\frac{\partial\mathcal{L}}{\partial(\partial_{0}\partial_i\partial_j\phi)}+\delta^{\alpha}_i\partial_j\partial_k\frac{\partial\mathcal{L}}{\partial(\partial_{i}\partial_j\partial_k\phi)}\nonumber\\  &  
=&\frac{a^2}{6}\delta^{\alpha}_0\partial_0\partial_0\partial_0\phi-\frac{a^2}{8}\delta^{\alpha}_{i}\partial_0\partial_0\partial_i\phi-\frac{a^2}{4}\delta^{\alpha}_{0}\partial_0\partial_i\partial_i\phi.
\end{eqnarray}
Now adding up these we get the terms in first bracket as $\Big(\partial^{\alpha}\phi+\frac{a^2}{3}\delta^{\alpha}_0\partial_0\partial_0\partial_0\phi-\frac{a^2}{2}\delta^{\alpha}_i\partial_i\partial_i\partial_0\phi-\frac{a^2}{2}\delta^{\alpha}_0\partial_i\partial_i\partial_0\phi+\frac{a^2}{4}\delta^{\alpha}_i\partial_i\partial^j\partial_j\phi\Big)\partial^{\mu}\phi$. Similarly, we calculate the remaining terms and deformed stress tensor is found to be
\begin{eqnarray}\label{st}
\hat{T}^{\alpha\mu}&=&\partial^{\alpha}\phi\partial^{\mu}\phi+\frac{a^2}{3}\delta^{\alpha}_{0}\partial_0\partial_0\partial_0\phi\partial^{\mu}\phi-\frac{a^2}{2}\delta^{\alpha}_{i}\partial_0\partial_0\partial_i\phi\partial^{\mu}\phi-\frac{a^2}{2}\delta^{\alpha}_{0}\partial_i\partial_0\partial_i\phi\partial^{\mu}\phi+\frac{a^2}{4}\delta^{\alpha}_{i}\partial_j\partial_i\partial_j\phi\partial^{\mu}\phi\nonumber\\& & +\frac{3a^2}{8}\delta^{\alpha}_{i}\partial_i\partial_0\phi\partial_0\partial^{\mu}\phi+\frac{3a^2}{8}\delta^{\alpha}_{0}\partial_0\partial_i\phi\partial_i\partial_{\mu}\phi-\frac{a^2}{4}\delta^{\alpha}_{i}\partial_i\partial_j\phi\partial_j\partial^{\mu}\phi-\frac{a^2}{6}\delta^{\alpha}_{0}\partial_0\partial_0\phi\partial_0\partial^{\mu}\phi+\frac{a^2}{8}\delta^{\alpha}_{0}\partial_i\partial_i\phi\partial_0\partial^{\mu}\phi\nonumber\\& & +\frac{a^2}{6}\delta^{\alpha}_{0}\partial_0\phi\partial_0\partial_0\partial^{\mu}\phi-\frac{a^2}{8}\delta^{\alpha}_{i}\partial_i\phi\partial_0\partial_0\partial^{\mu}\phi-\frac{a^2}{8}\delta^{\alpha}_{0}\partial_i\phi\partial_0\partial_i\partial^{\mu}\phi-\frac{a^2}{8}\delta^{\alpha}_{0}\partial_i\phi\partial_i\partial_0\partial^{\mu}\phi-\eta^{\alpha\mu}\mathcal{L}.
\end{eqnarray}
In the lim $a\rightarrow 0$, we recover the commutative stress tensor for the real scalar field. Note that, unlike the commutative case, here the $\kappa$-deformed stress tensor is not symmetric. Thus we see that a stress tensor which is symmetric in the commutative space-time need not be symmetric in non-commutative space-time. Next we symmetrise the Energy-Momentum tensor, and then re-express the deformed Energy-Momentum tensor as
\begin{equation}
 \hat{T}^{\alpha\mu}_{x,x'}=(A+B+C+D)\phi(x)\phi(x'),
\end{equation}
where
\begin{equation}
 A= \frac{1}{2}\bigg((\partial^{\alpha}\partial'^{\mu}+\partial^{\mu}\partial'^{\alpha})-\eta^{\alpha\mu}\Big(\partial_m\partial'_m-\partial_0\partial'_0-\frac{\lambda}{L}\delta(0)-\frac{\lambda'}{L}\delta(x-L)\Big)\bigg),
\end{equation}
\begin{eqnarray}
B&=&\frac{a^2}{2}\bigg(\frac{1}{3}\delta^{\alpha}_0\partial_0\partial_0\partial_0\partial'^{\mu}-\frac{1}{2}\delta^{\alpha}_i\partial_0\partial_0\partial_i\partial'^{\mu}-\frac{1}{2}\delta^{\alpha}_0\partial_i\partial_0\partial_i\partial'^{\mu}+\frac{1}{4}\delta^{\alpha}_i\partial_i\partial_i\partial_i\partial'^{\mu}+\frac{1}{6}\delta^{\alpha}_0\partial'_0\partial_0\partial_0\partial^{\mu}-\frac{1}{8}\delta^{\alpha}_i\partial'_i\partial_0\partial_0\partial^{\mu}\nonumber\\& &-\frac{1}{8}\delta^{\alpha}_i\partial'_i\partial_i\partial_0\partial^{\mu}-\frac{\eta^{\alpha\mu}}{2}\Big(\frac{1}{3}\partial'_0\partial_0\partial_0\partial_0-\frac{1}{4}\partial'_m\partial_m\partial_0\partial_0\Big)\bigg),
\end{eqnarray}
\begin{eqnarray}
 C&=&\frac{a^2}{2}\bigg(\frac{3}{8}\delta^{\alpha}_i\partial_i\partial_0\partial'_0\partial'^{\mu}+\frac{3}{8}\delta^{\alpha}_i\partial'_i\partial'_0\partial_0\partial^{\mu}\phi(x)+\frac{3}{8}\delta^{\alpha}_0\partial_i\partial_0\partial'_i\partial'^{\mu}+\frac{3}{8}\delta^{\alpha}_0\partial'_i\partial'_0\partial_i\partial^{\mu}-\frac{1}{4}\delta^{\alpha}_i\partial_i\partial_i\partial'_i\partial'^{\mu}\nonumber\\& &-\frac{1}{4}\delta^{\alpha}_i\partial'_i\partial'_i\partial_i\partial^{\mu}-\frac{1}{6}\delta^{\alpha}_0\partial_0\partial_0\partial'_0\partial'^{\mu}-\frac{1}{6}\delta^{\alpha}_0\partial'_0\partial'_0\partial_0\partial^{\mu}+\frac{1}{8}\delta^{\alpha}_0\partial_i\partial_i\partial'_0\partial'^{\mu}+\frac{1}{8}\delta^{\alpha}_0\partial'_i\partial'_i\partial_0\partial^{\mu}\nonumber\\& &-\frac{\eta^{\alpha\mu}}{4}\Big(\frac{1}{2}\partial'_m\partial'_m\partial_m\partial_m-\frac{1}{2}\partial_m\partial_0\partial'_m\partial'_0\Big)\bigg),
\end{eqnarray}
and
\begin{eqnarray}
    D&=&\frac{a^2}{2}\bigg(\frac{1}{3}\delta^{\alpha}_0\partial'_0\partial'_0\partial'_0\partial^{\mu}-\frac{1}{2}\delta^{\alpha}_i\partial'_0\partial'_0\partial'_i\partial^{\mu}-\frac{1}{2}\delta^{\alpha}_0\partial'_i\partial'_0\partial'_i\partial^{\mu}+\frac{1}{4}\delta^{\alpha}_i\partial'_i\partial'_i\partial'_i\partial^{\mu}+\frac{1}{6}\delta^{\alpha}_0\partial_0\partial'_0\partial'_0\partial'^{\mu}-\frac{1}{8}\delta^{\alpha}_i\partial_i\partial'_0\partial'_0\partial'^{\mu}\nonumber\\& &-\frac{1}{4}\delta^{\alpha}_0\partial_i\partial'_0\partial'_i\partial'^{\mu}-\frac{\eta^{\alpha\mu}}{2}\Big(\frac{1}{3}\partial_0\partial'_0\partial'_0\partial'_0-\frac{1}{4}\partial_m\partial'_m\partial'_0\partial'_0\Big)\bigg),
\end{eqnarray}
In the above equations we have used  $\partial_\mu=\frac{\partial}{\partial x^\mu}$ and $\partial'_\mu=\frac{\partial}{\partial x'^\mu}$

Note that B,C and D are dependent on $a^2$ and in the commutative limit, they all vanish identically.

We now re-express the Energy-Momentum tensor as,
\begin{equation}
 \hat{T}^{\alpha\mu}_{xx'}=\hat{O}^{\alpha\mu}(\partial,\partial')T(\phi(x),\phi(x')),
\end{equation}
where $\hat{O}^{\alpha\mu}(\partial,\partial')=(A+B+C+D)$. Thus the vacuum expectation value of the $\kappa$-deformed stress tensor is
\begin{equation}
 <\hat{T}^{\mu\nu}>=\hat{O}^{\mu\nu}(\partial,\partial')<T(\phi(x),\phi(x'))>=-i\hat{O}^{\mu\nu}(\partial,\partial')G_a(x,x')
\end{equation}
Here the subscript $a$ in $G_a$ is to emphasis that this Green's function has $a$ dependence.

In order to evaluate the time order product between the scalar fields, we need to evaluate $\phi(x)\phi(x')$, Using Eq.(\ref{p1}), we get,
\begin{eqnarray}\label{pr}
 \phi(x)\phi(x')&=&\Big(1+2a\alpha\Big)\phi_0(x)\phi_0(x')+a^2\Big(\alpha^2\phi_0(x)\phi_0(x')+\beta\big(\phi_0(x)\phi_2(x')+\phi_2(x)\phi_0(x')\big)\Big)\nonumber\\
&=&\Big(1+2a\alpha+a^2\alpha^2-a^2\frac{13}{12}w^2\Big)\phi_0(x)\phi_0(x'),
\end{eqnarray}
Note that this product of fields do have terms up to $a^2$ order due to $\kappa$-deformation.
Using this, the vacuum expectation value of  deformed stress tensor is given as
\begin{equation}
 <\hat{T}^{\mu\nu}>=-i\hat{O}^{\mu\nu}(\partial,\partial')\Big(1+2a\alpha+a^2\alpha^2-a^2\frac{13}{12}w^2\Big)G(x,x').
\end{equation}
The Green's function appearing here is the same as in the commutative case (i.e  $G(x,x')=T(\phi(x)\phi(x'))$).
From the above expression, we obtain the vacuum expectation value of $xx$ component of the deformed stress tensor as
\begin{eqnarray}
 <\hat{T}^{xx}>&=&-i\Big(1+2a\alpha+a^2\alpha^2-a^2\frac{13}{12}w^2\Big)\bigg(\frac{1}{2}\partial_x\partial'_{x}+\frac{1}{2}\partial_0\partial'_{0}-\frac{a^2}{12}\partial_0\partial'_0\partial'_0\partial'_0-\frac{a^2}{12}\partial'_0\partial_0\partial_0\partial_0-\frac{a^2}{4}\partial_0\partial_0\partial_x\partial'_x\nonumber\\& &-\frac{a^2}{4}\partial'_0\partial'_0\partial'_x\partial_x+\frac{a^2}{4}\partial_0\partial_x\partial'_x\partial'_0-\frac{a^2}{8}\partial_x\partial_x\partial'_x\partial'_x+\frac{a^2}{8}\partial_x\partial_x\partial_x\partial'_x+\frac{a^2}{8}\partial'_x\partial'_x\partial'_x\partial_x\bigg)G(x,x'),
\end{eqnarray}
Using the reduced Stress tensor $\hat{t}^{\mu\nu}$ defined by
\begin{equation}
 <\hat{T}^{\mu\nu}>=\int \frac{dw}{2\pi}\hat{t}^{\mu\nu},
\end{equation}
we calculate $\hat{t}_{xx}$ and express it in terms of the reduced Green's function. For the calculational simplicity we denote $\hat{t}_{xx}=\hat{t}^{(0)}_{xx}+\hat{t}^{(1)}_{xx}$, where $\hat{t}^{(0)}_{xx}$ and $\hat{t}^{(1)}_{xx}$ are given by
\begin{equation}\label{t1}
 \hat{t}^{(0)}_{xx}=\frac{1}{2i}\Big(1+2a\alpha+a^2\alpha^2-a^2\frac{13}{12}w^2\Big)\bigg\{w^2+\partial_x\partial'_x\bigg\}g(x,x')\Big|_{x=x'}
\end{equation}
and
\begin{equation}\label{t2}
  \hat{t}^{(1)}_{xx}=\frac{1}{2i}\Big(1+2a\alpha+a^2\alpha^2-a^2\frac{13}{12}w^2\Big)\bigg\{\frac{a^2}{3}w^4+\frac{3a^2}{2}w^2\partial_x\partial'_x+\frac{a^2}{4}\partial_x\partial_x\partial_x\partial'_x-\frac{a^2}{4}\partial_x\partial_x\partial'_x\partial'_x+\frac{a^2}{4}\partial_x\partial'_x\partial'_x\partial'_x\bigg\}g(x,x')\Big|_{x=x'} ,
\end{equation}
respectively.
Note that correction due to non-commutativity in $\hat{t}^{(0)}_{xx}$ comes due to the corrections appearing in the product of fields $\phi(x)\phi(x')$, (see Eq.(\ref{pr}).The correction in $\hat{t}^{(1)}_{xx}$ in Eq.(\ref{t2}) due to the $a$ dependent terms in Energy Momentum tensor(see Eq.(\ref{st})).
\subsection{$\kappa$-Deformed Casimir Force and Energy} 
Now we substitute the expression for reduced Green's function from Eq.(\ref{reg1}) into  Eq.(\ref{t1}) and after simplifying, we get
\begin{equation}
 \hat{t}^{(0)}_{xx}=\frac{1}{2i}\bigg\{-k+\frac{1}{2k\Delta}\frac{2\lambda\lambda'}{(2kL)^2}2w^2\bigg\}\Big(1+2a\alpha+a^2\alpha^2-a^2\frac{13}{12}w^2\Big).
\end{equation}
Using Eq.(\ref{del}), we get the $xx$ component of deformed stress tensor just to the left of the point $x=L$, valid up to $a^2$ term as
\begin{equation}
 \hat{t}^{(0)}_{xx}\Big|_{x=L^-}=-\frac{k}{2i}\bigg(1+\frac{2}{\big(\frac{2kL}{\lambda}+1\big)\big(\frac{2kL}{\lambda'}+1\big)e^{2kL}-1}\bigg)\Big(1+2a\alpha+a^2\alpha^2-a^2\frac{13}{12}w^2\Big).
\end{equation}
Similarly we substitute Eq.(\ref{reg2}) in Eq.(\ref{t1}) and after simplifying using Eq.({\ref{del}) we find the $xx$ component of the deformed stress tensor, just to the right of the point at $x=L$, valid up to $a^2$ term as
\begin{equation}
 \hat{t}^{(0)}_{xx}\Big|_{x=L^+}=-\frac{k}{2i}\Big(1+2a\alpha+a^2\alpha^2-a^2\frac{13}{12}w^2\Big).
\end{equation}
We adopt the same procedure to calculate $\hat{t}^{(1)}_{xx}$. First we substitute the expression for reduced Green's function from Eq.(\ref{reg1}) in Eq.(\ref{t2}) and after simplification, we get
\begin{multline}
 \hat{t}^{(1)}_{xx}=\frac{a^2k^4}{2i}\bigg\{\frac{1}{3}-\frac{9}{4}\bigg(\frac{1}{2k}-\frac{1}{2k\Delta}\frac{2\lambda\lambda'}{(2kL)^2}\bigg)-\frac{7}{4}\frac{1}{2k\Delta}\frac{\lambda'}{2kL}\Big(1+\frac{\lambda}{2kL}\Big)e^{2kx}\\-\frac{7}{4}\frac{1}{2k\Delta}\frac{\lambda}{2kL}\Big(1+\frac{\lambda'}{2kL}\Big)e^{-2k(x-L)}\bigg\}\Big(1+2a\alpha+a^2\alpha^2-a^2\frac{13}{12}w^2\Big)
\end{multline}
Thus the $xx$ component of the deformed stress tensor, just to the left of the point $x=L$, valid up to $a^2$ term is
\begin{multline}
 \hat{t}^{(1)}_{xx}\Big|_{x=L^-}=\frac{a^2k^4}{2i}\bigg\{\frac{1}{3}-\frac{9}{4}\bigg(\frac{1}{2k}-\frac{1}{2k\Delta}\frac{2\lambda\lambda'}{(2kL)^2}\bigg)-\frac{7}{4}\frac{1}{2k\Delta}\frac{\lambda'}{2kL}\Big(1+\frac{\lambda}{2kL}\Big)e^{2kL}-\frac{7}{4}\frac{1}{2k\Delta}\frac{\lambda}{2kL}\Big(1+\frac{\lambda'}{2kL}\Big)\bigg\}
\end{multline}
Next we substitute the expression for reduced Green's function given in Eq.(\ref{reg2}) in Eq.(\ref{t2}) and after simplification, we arrive at
\begin{multline}
 \hat{t}^{(1)}_{xx}=\frac{a^2k^4}{2i}\bigg\{\frac{1}{3}-\frac{9}{4}\frac{1}{2k}-\frac{7}{4}\frac{1}{2k\Delta}\bigg(\frac{\lambda}{2kL}\Big(1-\frac{\lambda'}{2kL}\Big)+\frac{\lambda'}{2kL}\Big(1+\frac{\lambda}{2kL}\Big)e^{2kL}\bigg)e^{-2k(x-L)}\bigg\}\Big(1+2a\alpha+a^2\alpha^2-a^2\frac{13}{12}w^2\Big)
\end{multline}
Using this, the $xx$ component of the deformed stress tensor, just to the right of the point $x=L$, valid up to $a^2$ term is
\begin{multline}
\hat{t}^{(1)}_{xx}\Big|_{x=L^+}=\frac{a^2k^4}{2i}\bigg\{\frac{1}{3}-\frac{9}{4}\frac{1}{2k}-\frac{7}{4}\frac{1}{2k\Delta}\bigg(\frac{\lambda}{2kL}\Big(1-\frac{\lambda'}{2kL}\Big)+\frac{\lambda'}{2kL}\Big(1+\frac{\lambda}{2kL}\Big)e^{2kL}\bigg)\bigg\}
\end{multline}
The $\kappa$-deformed Casimir force acting on the point $x=L$ due to the quantum fluctuations in the deformed scalar field is given by
\begin{eqnarray}
 \hat{F}&=&<\hat{T}^{xx}>\bigg| _{x=L^-}-<\hat{T}^{xx}>\bigg|_{x=L^+}\nonumber\\
&=&\int \frac{dw}{2\pi}\Big(\hat{t}^{(0)}_{xx}\Big|_{x=L^-}-\hat{t}^{(0)}_{xx}\Big|_{x=L^+}+\hat{t}^{(1)}_{xx}\Big|_{x=L^-}-\hat{t}^{(1)}_{xx}\Big|_{x=L^+}\Big)\nonumber\\&=&\hat{F}^{(0)}+\hat{F}^{(1)}.
\end{eqnarray}
Where
\begin{equation}
\begin{split}
 \hat{F}^{(0)}=&\Big(1+2a\alpha+a^2\alpha^2\Big)\int\frac{dw}{2\pi}\frac{ik}{\Big(\frac{2kL}{\lambda}+1\Big)\Big(\frac{2kL}{\lambda'}+1\Big)e^{2kL}-1}-a^2\frac{13}{12}\int\frac{dw}{2\pi}w^2\frac{ik}{\Big(\frac{2kL}{\lambda}+1\Big)\Big(\frac{2kL}{\lambda'}+1\Big)e^{2kL}-1}\\
=&\Big(1+2a\alpha+a^2\alpha^2\Big)\int\frac{dk}{2\pi}\frac{k}{\Big(\frac{2kL}{\lambda}+1\Big)\Big(\frac{2kL}{\lambda'}+1\Big)e^{2kL}-1}-a^2\frac{13}{12}\int\frac{dk}{2\pi}\frac{k^3}{\Big(\frac{2kL}{\lambda}+1\Big)\Big(\frac{2kL}{\lambda'}+1\Big)e^{2kL}-1}\\
=&-\Big(1+2a\alpha+a^2\alpha^2\Big)\frac{1}{4\pi L^2}\int_0^{\infty}dy\frac{y}{\Big(\frac{y}{\lambda}+1\Big)\Big(\frac{y}{\lambda'}+1\Big)e^y-1}-a^2\frac{13}{384\pi L^4}\int_0^{\infty}dy\frac{y^3}{\Big(\frac{y}{\lambda}+1\Big)\Big(\frac{y}{\lambda'}+1\Big)e^y-1}
\end{split}
\end{equation}
and
\begin{eqnarray}
\hat{F}^{(1)}&=&\int\frac{dw}{2\pi}\frac{a^2k^4}{2i}\frac{\lambda\lambda'}{2k\Delta(2kL)^2}
=\int\frac{dw}{2\pi}\frac{a^2k^3}{4i}\frac{1}{\Big(\frac{2kL}{\lambda}+1\Big)\Big(\frac{2kL}{\lambda'}+1\Big)e^{2kL}-1}\nonumber\\
&=&\frac{a^2}{128\pi L^4}\int_0^{\infty} dy \frac{y^3}{\Big(\frac{y}{\lambda}+1\Big)\Big(\frac{y}{\lambda'}+1\Big)e^y-1}.
\end{eqnarray}
Thus the deformed Casimir force is obtained by adding $\hat{F}^{(0)}$ and $\hat{F}^{(1)}$ as
\begin{equation}\label{cf1}
 \hat{F}_{\lambda,\lambda'\rightarrow \infty}=-\Big(1+2a\alpha+a^2\alpha^2\Big)\frac{\pi}{24L^2}-a^2\frac{7\pi^3}{11520L^4}.
\end{equation}
Hence the deformed Casimir energy is obtained as
\begin{equation}\label{en}
 \hat{E}_{\lambda,\lambda'\rightarrow \infty}=-\int_0^{\infty} dx\hat{F}_{\lambda,\lambda'\rightarrow\infty}=-\Big(1+2a\alpha+a^2\alpha^2\Big)\frac{\pi}{24L}-a^2\frac{7\pi^3}{34560L^3}.
\end{equation}
In $3+1$ dimension, we evaluate the deformed Casimir pressure between $\delta$ function planes lying at $x=0$ and $x=L$ in three spatial dimensions. Here the Green's function become
\begin{equation}
 G(x,x')=\int \frac{dw}{2\pi}e^{iw(t-t')}\int \frac{d^3k}{(2\pi)^3}e^{i\vec{k}\cdot(\vec{r}-\vec{r}')}g(x,x';k'),
\end{equation}
where $k'^2=k^2-w^2$. Then $g(x,x',k')$ has the same form as in Eq.(\ref{reg1}),Eq.(\ref{reg2}) and Eq.(\ref{reg3}), respectively. So that the deformed Casimir pressure, $\hat{P}=\frac{\hat{F}}{A}$ is given as
\begin{eqnarray}\label{cf2}
 \hat{P}&=&-\Big(1+2a\alpha+a^2\alpha^2\Big)\int\frac{dw}{2\pi^2}\frac{-ik^3}{\Big(\frac{2kL}{\lambda}+1\Big)\Big(\frac{2kL}{\lambda'}+1\Big)e^{2kL}-1}-a^2\frac{7}{24}\int\frac{dw}{2\pi^2}\frac{-ik^5}{\Big(\frac{2kL}{\lambda}+1\Big)\Big(\frac{2kL}{\lambda'}+1\Big)e^{2kL}-1}\nonumber\\
&=&-\Big(1+2a\alpha+a^2\alpha^2\Big)\frac{1}{32\pi^2 L^4}\int_0^{\infty}dy\frac{y^3}{\Big(\frac{y}{\lambda}+1\Big)\Big(\frac{y}{\lambda'}+1\Big)e^y-1}-a^2\frac{7}{3072\pi^2 L^6}\int_0^{\infty}dy\frac{y^5}{\Big(\frac{y}{\lambda}+1\Big)\Big(\frac{y}{\lambda'}+1\Big)e^y-1}\nonumber\\
&=&-\Big(1+2a\alpha+a^2\alpha^2\Big)\frac{\pi^2}{480 L^4}-a^2\frac{7}{3072\pi^2 L^6}5!\zeta(6)
\end{eqnarray}
and Casimir Energy,
\begin{eqnarray}
\hat{E}&=&-\Big(1+2a\alpha+a^2\alpha^2\Big)\frac{\pi^2}{120 L^3}-a^2\frac{7}{512\pi^2 L^5}5!\zeta(6).
\end{eqnarray}
where $A$ is surface area on the plate and $\zeta(n)$ is zeta function for $n=6$.

Note that in $1+1$ dimension, for deformed Casimir Force, the $\kappa$-deformed correction terms scale as $\frac{1}{L^2}$ and $\frac{1}{L^4}$ (see Eq.(\ref{cf1})). Here corrections with $\frac{1}{L^2}$ terms are coming due to correction in $\phi(x)\phi(x')$ (see Eq.(\ref{pr})). Modification with$\frac{1}{L^4}$ terms appear due to corrections in the product of fields $\phi(x)\phi(x')$ (see Eq.(\ref{pr})) and due to to $\kappa$-deformed correction of the Energy-Momentum tensor (see Eq.(\ref{st})). In $3+1$ dimensions, due to $\kappa$-deformation, Casimir force comes with terms, which has dependency of $\frac{1}{L^4}$ and $\frac{1}{L^6}$ (see Eq.(\ref{cf2})). Here corrections in $\frac{1}{L^4}$ related terms, are only because of corrections in product of fields $\phi(x)\phi(x')$ (see Eq.(\ref{pr})) and correction in $\frac{1}{L^6}$ terms are due to correction in Energy-Momentum tensors, due $\kappa$-deformation (see Eq.(\ref{st})).
\subsection{Bound on $\kappa$-Deformation parameter $a$}
In \cite{sed}, the Casimir force gradient is measured to be $8\times 10^{-4}N/m^2$ with an error less than $1\%$, for parallel plates separated by a distance ($L$) of $10\mu m$. Using the expression for deformed Casimir force given in Eq.(\ref{cf2}) for $3+1$ dimensions and by setting the numerical value of $\alpha$ to unity, we find the bound on the deformation parameter to be $a<10^{-23}$m.
\section{Conclusions}
In this work we have studied the vacuum fluctuations in $\kappa$-deformed space-time by analysing the Casimir effect between two parallel plates in $1+1$ and generalised these results to $3+1$ dimensions. We have calculated $\kappa$-deformed corrections to Casimir force and energy between two parallel plates. In $1+1$ dimensions, we have seen that between two parallel plates, in $\kappa$-deformed space-time, Casimir force and Casimir energy both pick up three correction terms. In these three correction terms, one depends on $a$ and remaining two terms are $a^2$ dependent. The $a$ dependent correction term varies as $\frac{1}{L^2}$ in Casimir force and scale as $\frac{1}{L}$ in Casimir energy. Similarly one of the $a^2$ dependent correction term in deformed Casimir force varies as $\frac{1}{L^2}$ and other $a^2$ dependent term which scale as $\frac{1}{L^4}$. This second term is due to correction introduced by $\kappa$-deformed energy momentum tensor. These former terms are due to the correction introduced by the product of fields in $\kappa$-space-time (see Eq.(\ref{pr})). Modified $a^2$ dependent terms of Casimir energy varies as $\frac{1}{L}$ and $\frac{1}{L^3}$. In $3+1$ dimensions, we  have evaluated deformed Casimir force, which has correction terms that scale as $\frac{1}{L^4}$ and $\frac{1}{L^6}$. Unlike \cite{1112.2924} here the corrections to the Casimir energy scales in two ways, i.e, as $L^{-3}$ and $L^{-5}$, respectively. All the correction term appearing in the Casimir force are attractive in nature. Note that we have obtained the corrections to Casimir energy that scale as $L^{-5}$ in addition to terms that scale as $L^{-3}$. This should be contrasted with the result of \cite{npb127}, where the correction was scaling as $L^{-3}$, only. 

In commutative case, to calculate Casimir force, Dirichlet boundary condition has been applied. As defining boundary condition in Non-commutative space-time is not trivial, here we started with $\kappa$-deformed Klein-Gordon Lagrangian, (which is invariant under the action of undeformed $\kappa$-Poincare algebra) which is expressed in terms of fields defined in commutative space-time. This allowed us to use the same calculational tools with the same boundary condition as commutative case. Comparing correction terms with experimental results in $3+1$ dimension, we have calculated upper bound on deformation parameter $a$ as $10^{-23}$ m. 

Casimir effect due to electromagnetic field was studied and this field was handled using its equivalence to two scalar modes. In these studies, Dirichlet boundary condition was imposed on one mode and for the other mode, Neumann boundary condition was imposed, which was introduced by hand and did not come out directly from Lagrangian. But in \cite{kvs}, the boundary conditions were derived and this method was developed to deal electromagnetic field without considering individual modes, separately. Generalising this approach to non-commutative space-time is in progress.

\section*{Appendix}
We consider a higher order derivative field theory such that its Lagrangian possess the derivative terms of up to fourth order. i.e, $\mathcal{L}(\phi,\partial_{\mu}\phi,\partial_{\mu}\partial_{\nu}\phi,\partial_{\mu}\partial_{\nu}\partial_{\lambda}\phi)$. Now we vary the action for an infinitesimal change in the space-time coordinate given as $x'^{\mu}=x^{\mu}+\epsilon\delta x^{\mu}$
\begin{multline}\nonumber
 \delta S=\int d^4x'\mathcal{L'}(\phi',\partial_{\mu}\phi',\partial_{\mu}\partial_{\nu}\phi',\partial_{\mu}\partial_{\nu}\partial_{\lambda}\phi')-\int d^4x\mathcal{L}(\phi,\partial_{\mu}\phi,\partial_{\mu}\partial_{\nu}\phi,\partial_{\mu}\partial_{\nu}\partial_{\lambda}\phi)\\
=\int d^4x\delta\mathcal{L}+\epsilon\int dx^4(\partial_{\mu}\delta x^{\mu})\mathcal{L}\\
=\int d^4x\bigg[\frac{\partial\mathcal{L}}{\partial\phi}-\partial_{\mu}\frac{\partial\mathcal{L}}{\partial(\partial_{\mu}\phi)}+\partial_{\mu}\partial_{\nu}\frac{\partial\mathcal{L}}{\partial(\partial_{\mu}\partial_{\nu}\phi)}-\partial_{\mu}\partial_{\nu}\partial_{\lambda}\frac{\partial\mathcal{L}}{\partial(\partial_{\mu}\partial_{\nu}\partial_{\lambda}\phi)}\bigg]\delta\phi\\
+\int d^4x\partial_{\mu}\bigg[\epsilon\mathcal{L}\delta x^{\mu}+\bigg\{\frac{\partial\mathcal{L}}{\partial(\partial_{\mu}\phi)}-\partial_{\nu}\frac{\partial\mathcal{L}}{\partial(\partial_{\mu}\partial_{\nu}\phi)}+\partial_{\nu}\partial_{\lambda}\frac{\partial\mathcal{L}}{\partial(\partial_{\mu}\partial_{\nu}\partial_{\lambda}\phi)}\bigg\}\delta\phi\\
+\bigg\{\frac{\partial\mathcal{L}}{\partial(\partial_{\mu}\partial_{\nu}\phi)}-\partial_{\lambda}\frac{\partial\mathcal{L}}{\partial(\partial_{\mu}\partial_{\nu}\partial_{\lambda}\phi)}\bigg\}\delta(\partial_{\nu}\phi)
+\bigg\{\frac{\partial\mathcal{L}}{\partial(\partial_{\mu}\partial_{\nu}\partial_{\lambda}\phi)}\bigg\}\delta(\partial_{\nu}\partial_{\lambda}\phi)
\end{multline}
The terms in the first square bracket vanishes as it represents the Euler-Lagrange equation for the fourth order derivative field and using the relation $\delta\phi=\Delta\phi-\delta x^{\mu}\partial_{\mu}\phi$ we reduce the above equation to
\begin{multline}\nonumber
\delta S=\int d\sigma^{\mu}\Bigg[\bigg\{\frac{\partial\mathcal{L}}{\partial(\partial_{\mu}\phi)}-\partial_{\nu}\frac{\partial\mathcal{L}}{\partial(\partial_{\mu}\partial_{\nu}\phi)}+\partial_{\nu}\partial_{\lambda}\frac{\partial\mathcal{L}}{\partial(\partial_{\mu}\partial_{\nu}\partial_{\lambda}\phi)}\bigg\}\Delta\phi\\
+\bigg\{\frac{\partial\mathcal{L}}{\partial(\partial_{\mu}\partial_{\nu}\phi)}-\partial_{\lambda}\frac{\partial\mathcal{L}}{\partial(\partial_{\mu}\partial_{\nu}\partial_{\lambda}\phi)}\bigg\}\Delta(\partial_{\nu}\phi)
+\bigg\{\frac{\partial\mathcal{L}}{\partial(\partial_{\mu}\partial_{\nu}\partial_{\lambda}\phi)}\bigg\}\Delta(\partial_{\nu}\partial_{\lambda}\phi)\\
-\epsilon\bigg[\bigg\{\frac{\partial\mathcal{L}}{\partial(\partial_{\mu}\phi)}-\partial_{\nu}\frac{\partial\mathcal{L}}{\partial(\partial_{\mu}\partial_{\nu}\phi)}+\partial_{\nu}\partial_{\lambda}\frac{\partial\mathcal{L}}{\partial(\partial_{\mu}\partial_{\nu}\partial_{\lambda}\phi)}\bigg\}\partial^{\alpha}\phi\\
+\bigg\{\frac{\partial\mathcal{L}}{\partial(\partial_{\mu}\partial_{\nu}\phi)}-\partial_{\lambda}\frac{\partial\mathcal{L}}{\partial(\partial_{\mu}\partial_{\nu}\partial_{\lambda}\phi)}\bigg\}\partial_{\nu}\partial^{\alpha}\phi
+\bigg\{\frac{\partial\mathcal{L}}{\partial(\partial_{\mu}\partial_{\nu}\partial_{\lambda}\phi)}\bigg\}\partial_{\nu}\partial_{\lambda}\partial^{\alpha}\phi
-\eta^{\mu\alpha}\mathcal{L}\bigg]\delta x_{\alpha}\Bigg]
\end{multline}
Hence the stress tensor for a fourth order derivative field theory is obtained to be
\begin{multline}\nonumber
  T^{\mu\alpha}=\bigg\{\frac{\partial\mathcal{L}}{\partial(\partial_{\mu}\phi)}-\partial_{\nu}\frac{\partial\mathcal{L}}{\partial(\partial_{\mu}\partial_{\nu}\phi)}+\partial_{\nu}\partial_{\lambda}\frac{\partial\mathcal{L}}{\partial(\partial_{\mu}\partial_{\nu}\partial_{\lambda}\phi)}\bigg\}\partial^{\alpha}\phi\\
+\bigg\{\frac{\partial\mathcal{L}}{\partial(\partial_{\mu}\partial_{\nu}\phi)}-\partial_{\lambda}\frac{\partial\mathcal{L}}{\partial(\partial_{\mu}\partial_{\nu}\partial_{\lambda}\phi)}\bigg\}\partial_{\nu}\partial^{\alpha}\phi
+\bigg\{\frac{\partial\mathcal{L}}{\partial(\partial_{\mu}\partial_{\nu}\partial_{\lambda}\phi)}\bigg\}\partial_{\nu}\partial_{\lambda}\partial^{\alpha}\phi-\eta^{\mu\alpha}\mathcal{L}
\end{multline}
\subsection*{\bf Acknowledgments}
We thank K V Shajesh for useful discussions and suggestions. Two of us (EH and SKP) thank SERB, Govt. of India, for support through EMR/2015/000622. VR thanks Govt. of India, for support through DST-INSPIRE/IF170622.

\end{document}